\documentstyle[12pt]{article}
\def\beq{\begin{equation}}
\def\eeq{\end{equation}}
\def\bea{\begin{eqnarray}}
\def\eea{\end{eqnarray}}

\def\vel{\left|}
\def\ver{\right|}

\def\ba{\begin{array}}
\def\ea{\end{array}}
\def\ds{\displaystyle}

\def\0{\c{s}}
\def\,{\"{U}}
\def\6{\.{I}}   

\begin{document}

\title{ {\Large {\bf 
The theory of transvers Nernst--Ettingshausen effect and thermopower of
heated (hot) charge carriers in nondegenerate semiconductors in the 
case of high anisotropy of phonons distribution function } } }

\author{\vspace{1cm}\\
{\large T. M. Gassym$^{a,b,}$\footnote{e-mail:~
gassymt@newton.physics.metu.edu.tr}~~ and
\hskip 0.3 cm
\vspace{0.5cm}
H. A. Hasanov$^{a,}$}\\
\hskip 0.4 cm
$^a${\small Institute of Physics of Azerbaijan National Academy 
of Sciences, Baku 370143, Azerbaijan}\\
\hskip 0.5 cm
$^b${\small Physics Department, Middle East Technical University
06531 Ankara, Turkey} }

\date{}

\begin{titlepage}
\maketitle   
\thispagestyle{empty}

\begin{abstract}
The transverse Nernst--Ettingshausen (NE) effect and thermopower of hot
charge carriers in nondegenerate impure semiconductors placed at high
electric and nonquantized magnetic fields in the nondiffusion approximation 
is studied. Arbitrary heating and mutual drag of electrons and long
wavelength phonons interacting with electrons are considered. 

The spectrum of electrons is assumed to be strong nonparabolic in Kane
two--band approximation.

The case when the electron concentration is high and frequent
interelectronic collisions lead to the equilibrium symmetric part 
of the electron distribution function with effective electron 
temperature is considered.
\end{abstract}
\end{titlepage}

\section{Introduction}
The theoretical and experimental interest in thermoelectric power
in bulk and recent low dimensional systems has been
intensified\cite{1}--\cite{12}.

A relatively long survey of literature and some common misunderstandings
in the field of thermoelectric power and NE effect under different
transport conditions\cite{3,7,8,9},\cite{15}--\cite{23} is given in our
recent paper\cite{Kane}.

X. L. Lei\cite{11} showed that the diffusion component of thermoelectric
power may be negative within a low enough lattice temperature range at
high electric field while the phonon drag component is still positive.
It should be noted that such a result is obtained by Babaev and
Gassymov\cite{20}. In that paper, the thermoelectric power and
transverse Nernst--Ettingshausen (NE) effect in semiconductors at high  
electric and nonquantizing magnetic fields are theoretically investigated
by solving the coupled system of kinetic equations for electrons and   
phonons by taking into account the heating of electrons and phonons, and
the phonon drag. It is shown that when the temperature gradient of hot
electrons is produced only by the lattice temperature gradient,
$\nabla E=0, \nabla T_e= {\ds \frac{\partial T_e}{\partial T}} \nabla T$,
the electronic parts of the thermoelectric and NE fields reverse their
sign. In the case of heated phonons, when $T_p=T_e \gg T$ both the
electronic and phonon parts of the thermoelectric and thermomagnetic
fields reverse their sign for all cases considered. Here $T_e$, $T_p$ 
and $T$ are the temperature of electrons, phonons and lattice in energy
units, respectively.

The interest in the study of thermoelectric power and NE effect in II--VI
semiconductors has been increased\cite{31}--\cite{34}. The earlier 
investigations of the magnetic field dependence of the longitudinal NE
effect in HgSe\cite{35,36} and lead chalcogenides\cite{37,38} in the 
region of comparatively high temperatures $(T \geq 77 K)$ demonstrated 
that the thermo emf exhibits saturation in the classical region of   
strong magnetic fields $(H)$ irrespective of the dominant mechanism of 
charge carrier scattering in the conduction band. However, the
measurements of the longitudinal NE effect in iron--doped HgSe samples
at low temperatures $(20 \leq T \leq 60 K)$ revealed the presence of 
a maxima in the plot of $\Delta\alpha(H)=\mid \alpha(H)-\alpha(0)\mid$.  
$\Delta\alpha(H)$ first increases quadratically with increasing $H$
for $\omega_H \tau < 1$ (here ${\ds \omega_H=\frac{eH}{mc}}$ is the   
cyclotron frequency and $\tau$ is the electron relaxation time), then
passes through a maximum for some $H=H_m$, and finally decreases as the
field increases further. Another unusual fact is the sign reversal of
the transverse NE coefficient $Q_{\perp}(H)$ with increasing magnetic  
field in the range $\omega_H \tau > 1$\cite{33,34}. The experiments in
Ga--doped HgSe showed that at low temperatures the NE coefficients
change sign with increasing Ga~ concentration or the applied magnetic
field strength. The unusual features of the NE effect observed in HgSe   
crystals may be attributed to the effect of mutual drag which can  
experimentally be detected in semiconductors with high concentration  
of conduction electrons\cite{39}.

In the absence of external magnetic field, the thermopower of hot
electrons taking into account the heating of phonons and the thermal 
drag is considered in \cite{19}. The consideration is made for the
deformation
potential of interaction between electrons and phonons. The thermopower
and transverse NE effect of hot electrons in nondegenerate semiconductors
are studied in \cite{41} without taking into account the effect of phonon
drag and their heating; and in \cite{20} by taking into account    
the thermal drag only in transverse magnetic field. Neither in these  
studies the mutual drag of charge carriers and phonons are considered.

A consistent microscopic theory of transport phenomena in semiconductors
and semimetals in high external electric and magnetic fields with due  
regard for the heating of charge carriers and phonons, their thermal and  
mutual drags, and the possible phonon generation by the drift charge   
carriers must be based on the solution of coupled system of kinetic
equations for charge carriers and phonons. Such a problem is formulated
and solved for the first time by Gassymov\cite{28}, see also reference
\cite{27}. In the statement of the problem, it should be noted that the
traditional approximation of small anisotropy of phonon distribution
function (so--called ``diffusion approximation") is applicable to phonons
whose drift velocities ($u$) is much smaller than the sound velocity
($s_0$) in crystal. In the presence of external electric and magnetic  
fields, this condition obviously is not fulfilled. This violation shows up
particularly in several ways under the acoustical instability conditions
$(u \geq s_0)$. Actually, both spherically symmetric, $N_s(q)$, and
antisymmetric, $N_a(q)$, parts of the phonon distribution function as   
well as ${\ds \frac{N_a(q)}{N_s(q)}}$ grow as $u$ increases. Indeed,  
${\ds \frac{N_a(q)}{N_s(q)} \rightarrow 1}$ as $u \rightarrow s_0$, and
${\ds \frac{N_a(q)}{N_s(q)} \gg 1}$ when $u \gg s_0$. The general solution
of the Boltzmann equation for phonons shows that $N(q)$ is stationary for
$u<s_0$, and nonstationary for $u \geq s_0$. These results are obtained by
solving the nonstationary kinetic equation for phonons interacting with
charge carriers at high electric and arbitrary magnetic fields in the
nondiffusion approximation\cite{27,28,29}.

The theoretical investigation of any thermo and galvanomagnetic effects 
is usually based on solving the kinetic equations of electrons and phonons 
in so--called ``diffusion approximation (DA)". It is the approximation of
small anisotropy of distribution function (DF) of charge carriers and
phonons and it is applicable when the drift velocities of the carriers 
$v(\varepsilon)$ and phonons $u$ are much less than their thermal (chaotic) 
velocities $v_T$ and $s_0$ respectively. Here ${\ds v_T =(2 T/m_n)^{1/2}}$ 
is the thermal velocity of charge carriers, $m_n$ is the effective mass of
electrons at the bottom of the conduction band. It was shown in \cite{29}
at high external electric field under the conditions of strong mutual drag
at low temperatures and near the point of acoustic instability threshold
(AIT) for arbitrary temperatures DA is not applicable for the phonons. 
Because, the drift velocity of phonons under these conditions can be equal
or more than the velocity of sound $s_0$ and the condition $u \ll s_0$ for
DA is violated. On the other hand, under the same conditions the DA can
be applicable for electrons because in this case ${\ds v_T/s_0 \approx
10^2,10^3}$.

For the conservation of usual scheme of consideration, we separate the 
distribution function of phonons, $N({\bf q})$, into its symmetric, 
$N_s({\bf q})$, and antisymmetric, $N_a({\bf q})$, parts 

\beq
N({\bf q})=N_s({\bf q})+N_a({\bf q}).
\eeq
Note that in this case the relative values of $N_s({\bf q})$ and 
$N_a({\bf q})$ are arbitrary in contrast to DA when we assume that 
$N_a({\bf q}) \ll N_s({\bf q})$, which is equivalent to the assumption
$u \ll s_0$. Therefore, in the case considered DA is applicable for
electrons (charge carriers) and it is not applicable for phonons. In 
the light of foregoing discussion we will search the DF of electrons 
in the form:

\beq
f(\varepsilon,r)=f_0(\varepsilon,r)+f_1(\varepsilon,r),~~~~~
\mid f_1(\varepsilon,r) \mid \ll f_0(\varepsilon,r),
\eeq
where $f_0(\varepsilon,r)$ and $f_1(\varepsilon,r)$ are the 
spherically symmetric and antisymmetric parts of the distribution 
function of electrons, respectively. 

The present work is dealing with theoretical investigation of the
transverse Nernst--Ettingshausen (NE) effect and thermopower of hot 
charge carriers in nondegenerate impure semiconductors placed at high 
electric ${\bf E}$ and nonquantized magnetic fields ${\bf H}$ in 
nondiffusion approximation. The arbitrary heating and mutual drag of 
electrons and long wavelength (LW) phonons interacting with 
electrons are considered. The phonon part of the NE voltage differs from 
zero only for the nonparabolic spectrum of electrons in that the spectrum
of electrons is assumed to be strong nonparabolic in Kane two--band
approximation:

\beq
p(\varepsilon)=\mu \varepsilon^s,
\eeq
where $\mu=\sqrt{2 m_n}$, ${\ds s=\frac{1}{2}}$ for parabolic, and 
${\ds \mu=\sqrt{\frac{2m_n}{\varepsilon_g}}}$, $s=1$ for strong
nonparabolic spectrum of electrons with band gap $\varepsilon_g$. 
Consider the case of high concentration of electrons when 
the frequent interelectronic collisions lead to the equilibrium 
symmetric part of electrons distribution function with effective
temperature $T_e \gg T$

\bea
f_0(\varepsilon)={\ds \frac{2\pi^2\hbar^3n_0}{(2m_nT_e)^{\frac{3}{2}}
F_0(T_e)}}~exp \left(-\frac{\varepsilon}{T_e} \right), \\
\nonumber 
F_0(T_e)={\ds \frac{s\Gamma(3s)}{\sqrt 2}\Lambda^{\frac{3}{2}}
\ds \Theta_e^{3(s-\frac{1}{2})}}.
\eea
where ${\ds \Lambda= \frac{\mu^2~T^{2s-1}}{m_n}}$,~ ${\ds \Theta_e=
\frac{T_e}{T}}$ is the dimensionless temperature of electrons, $n_0$ is 
electron concentration and $\Gamma(x)$ is the Euler's gamma function. 

\section{The Main Equations and Their Solutions}
From the stationary kinetic equation of phonons in the presence of 
space nonhomogenity by taking into account Eqs. (1) and (2), we have 
the following system of equations for $N_a({\bf q})$ and 
$N_s({\bf q})$\cite{27,28}:

\bea
N_a({\bf q})+\frac{s_0}{\beta(q)}~ \frac{{\bf q}\nabla N_s(q)}{q}
=\frac{u}{s_0}N_s(q)\cos\gamma, \\
\nonumber
N_s({\bf q})+\frac{s_0}{\beta(q)}~\frac{{\bf q}\nabla N_a(q)}{q}
=\tilde{N}(q,\tilde{T})+\frac{u}{s_0}N_a(q)\cos\gamma,
\eea
where $\gamma$ is the angle between ${\bf q}$ and ${\bf u}$, $\beta(q)=
\beta_e(q)+\beta_{pb}(q)$ is the total collision frequency of phonons
with scattering centers, and $\beta_{pb}(q)=\beta_p(q)+\beta_b(q)$. The 
subindices of $\beta$ stand for the scattering of phonons by electrons 
($e$), SW phonons ($p$), and crystal boundaries ($b$). 
$\tilde{N}(q,\tilde{T})$ is given by

\bea
\tilde{N}(q,\tilde{T})=\frac{\beta_e(q)}{\beta(q)}N(q,T_e)+
\frac{\beta_p(q)}{\beta(q)}N(q,T_p)+\frac{\beta_b(q)}{\beta(q)}N(q,T), \\
\nonumber
\tilde{T}=\frac{\beta_e(q)}{\beta(q)}T_e+\frac{\beta_p(q)}{\beta(q)}T_p+
\frac{\beta_b(q)}{\beta(q)}T.
\eea
Here $N(q,T_e)$, $N(q,T_p)$ and $N(q,T)$ are the Planck distribution
functions with $T_e$, $T_p$ and $T$, respectively.

In the absence of space nonhomogenity, from Eq. (5) we have

\beq
N_s({\bf q})=\frac{\tilde{N}(q,\tilde{T})}{1-\left({\ds\frac{u}{s_0}}
\right)^2 \cos^2\gamma}, ~~~~~
N_a({\bf q})=\frac{\tilde{N}(q,\tilde{T})}{1-\left({\ds\frac{u}{s_0}}
\right)^2 \cos^2\gamma}\frac{u}{s_0}\cos\gamma.
\eeq

From Eq. (7), for the total distribution function of phonons we obtain

\beq
N({\bf q})=N_s({\bf q})+N_a({\bf q})=\frac{\tilde{N}(q,\tilde{T})}
{1-{\ds \frac{u}{s_0}}\cos\gamma}.
\eeq
The Eqs. (6)--(8) were obtained for the first time in \cite{28}. 

In the presence of space nonhomogenity, the symmetric part of the
distribution function $N_s({\bf q})$ is defined by Eq. (7), but 
the anisotropic (antisymmetric) part has the form

\beq
N_a({\bf q})=\frac{u}{s_0}N_s({\bf q})\cos\gamma-
\frac{s_0}{\beta(q)}\frac{{\bf q} \nabla N_s(q)}{q}.
\eeq

Choosing ${\ds {\bf f_1}(\varepsilon)= p {\bf v}(\varepsilon)
\left(-\frac{\partial f_0(\varepsilon)}{\partial \varepsilon}\right)}$, 
and substituting Eqs. (1), (2), (7) and (9) into stationary kinetic 
equation of electrons, the drift velocity of electrons is found to be
\bea
{\bf v}(\varepsilon)-{\ds \frac{\omega_H(\varepsilon)}
{\nu(\varepsilon,u)}}\left[{\bf h}{\bf v}(\varepsilon)\right]
+\frac{1}{m(\varepsilon)\nu(\varepsilon,u)}~~~~~~~~~~~~~~~~~~~~~~~~ \\
\nonumber
\left\{e {\bf E^{\prime}}+\left(\frac{\varepsilon-\zeta(T_e)}{T_e}\right)
\nabla T_e + {\ds \frac{4\pi m^2(\varepsilon)}{(2 \pi \hbar)^3
p^3(\varepsilon)}}\int_0^{2p}~ dq~ W_q~ q^2~ \delta(u) 
\frac{s_0}{\beta(q)}\hbar\omega_q \nabla N(q,T_p)-q{\bf
u}(q) N(q,T_p) \right\},
\eea
where, ${\bf h}={\ds \frac{{\bf H}}{H}}$, ${\bf E^{\prime}}={\bf E}+
{\bf E_T}+ {\ds \frac{1}{e}} \nabla\zeta(T_e)$, ${\bf E_T}$ is the
thermomagnetic field, $\zeta(T_e)$ is the chemical potential of hot
electrons, and $\nu(\varepsilon,u)=\nu_p (\varepsilon,u)+
\nu_i(\varepsilon,u)$ is the total collision frequency of electrons 
by scatterers. Note that the subscript $p$ stands for phonons and 
$i$ for impurity ions. As it is shown in \cite{28}

\beq
\nu_p(\varepsilon,u)=\nu_p(\varepsilon)\delta(u),~~~\delta(u)=
\frac{3s_0^2}{u^2}\left(\frac{s_0}{2u}\varphi(u)-1\right), ~~~
\varphi(u)=\ln \vel \frac{1+\left({\ds \frac{u}{s_0}} \right)} 
{1-\left({\ds \frac{u}{s_0}}\right)} \ver.
\eeq

Since the isotropic part of the distribution function of electrons
$f_0(\varepsilon)$ is assumed to be the equilibrium one with effective
electron temperature $T_e$, the kinetic equation for the
$f_0(\varepsilon)$ will be used as a balance equation for the
definition of $T_e$ as a function of $E$ and $H$ later. In other words,
the kinetic equation for the $f_0(\varepsilon)$ is a single equation 
which connects $T_e$ with the amplitude of external electric and magnetic 
fields. For further calculations let represent $\nu(\varepsilon)$ and 
$\beta(q)$ in the form:

\bea
\nu(\varepsilon)=\nu(T)\left(1+2~\frac{\varepsilon}{\varepsilon_g}
\right)\left(1+\frac{\varepsilon}{\varepsilon_g}\right)^{-r}
\left(\frac{\varepsilon}{T}\right)^{-r},~~~
\nu(T_p)=\nu_0(T)\left(\frac{T_p}{T}\right)^{\ell}, \\
\nonumber
\beta(q)=\beta(\varepsilon_g,T)\left(\frac{q s}{T}\right)^{\kappa}~ 
\Theta^{-n},~~~\beta(\varepsilon_g,T)=\beta(T)\left(\frac{T}
{\varepsilon_g}\right)^{n(s-1/2)}
\eea
here $\kappa=0,1,t$, in the case of scattering of phonons by the boundaries 
of the specimen for the energy and momentum transfer, by SW phonons and
electrons, respectively. $n=2-s$ for the scattering of LW phonons by 
electrons and $n=0$ for other cases; $r$ is the parameter for
the scattering of electrons for the momentum transfer and $W_q=W_0~q^t$
is a constant of interaction with $t=1$ for deformation, and $t=-1$ for
piezoelectric interaction of electrons with acoustical phonons.

By using the solution of Eq. (10) for $\bf {V}(\varepsilon)$, we can 
calculate the current as follows:

\bea
{\bf J}=\sigma_{11}{\bf E^{\prime}}+\sigma_{12}\left[{\bf h}
{\bf E^{\prime}}\right]+\sigma_{13}{\bf h}\left[{\bf h}{\bf E^{\prime}}
\right]+\beta_{11}^e \nabla T_e+\beta_{12}^e \left[{\bf h} \nabla T_e
\right]+\beta_{13}^e {\bf h}\left[{\bf h}\nabla T_e \right] \\
\nonumber +{\ds \beta_{11}^p \nabla T_p+\beta_{12}^p \left[{\bf h}\nabla
T_ph \right]+\beta_{12}^p{\bf h}\left[{\bf h}\nabla T_p\right]};
\eea
where, 

\bea
\sigma_{1i}=\int_0^{\infty}d\varepsilon~\Psi(\varepsilon)
a_i(\varepsilon),~~~\beta_{1i}^e=\frac{1}{e}\int_0^{\infty}
d\varepsilon~a_i(\varepsilon)\left(\frac{\varepsilon-
\zeta(T_e)}{T_e}\right), \\
\nonumber
\beta_{1i}^p={\ds \frac{4\pi s_0\delta(u)}{(2\pi \hbar)^3}}
\int_0^{\infty}\frac{d\varepsilon~\Psi(\varepsilon)
a_i(\varepsilon)m^2(\varepsilon)}{p^3(\varepsilon)}\int_0^{2p} 
dq~\frac{W_q~ \hbar\omega_q}{\beta_q}q^2 \vel 
\frac{\nabla N(q,T_p)}{\nabla T_p} \ver,
\eea

\bea
\Psi(\varepsilon)=\frac{\nu(\varepsilon,u)~p^3(\varepsilon)
\left({\ds -\frac{\partial f_0(\varepsilon)}{\partial
\varepsilon}}\right)}{{\ds 3 \pi^2 \hbar^3 (1-\gamma)(1+\Omega^2)
m(\varepsilon)\left[\omega_H^2(\varepsilon)+
\nu^2(\varepsilon,u) \right]}},~~~a_1(\varepsilon)=\left(1-
\frac{\Omega~ \omega_H(\varepsilon)}{\nu(\varepsilon,u)}\right), \\
\nonumber 
a_2(\varepsilon)=\Omega+\frac{\omega_H(\varepsilon)}
{\nu(\varepsilon,u)},~~~a_3(\varepsilon)=\left(\frac{1+\Omega^2}
{1-\Gamma_0}\right)\left(1+\frac{\omega_H^2(\varepsilon)}
{\nu^2(\varepsilon,u)}\right)+\left(\Omega~
\frac{\omega_H(\varepsilon)}{\nu(\varepsilon,u)}-1\right),
\eea

\bea
\gamma=\frac{1}{a_4}\int_{\varepsilon(q/2)}^{\infty} 
d\varepsilon~\lambda(\varepsilon),~~~
a_4=\int_{\varepsilon(q/2)}^{\infty} d\varepsilon~m^2(\varepsilon)
\left(-\frac{\partial f_0(\varepsilon)}{\partial\varepsilon}\right), \\
\nonumber
\Omega={\ds \frac{1}{a_4(1-\gamma)}}\int_{\varepsilon(q/2)}^{\infty}
d\varepsilon~\lambda(\varepsilon)~\frac{\omega_H(\varepsilon)}
{\nu(\varepsilon,u)} 
\eea

\bea
\Gamma_0=\frac{1}{a_4(1-\gamma)}
\int_{\varepsilon(q/2)}^{\infty} d\varepsilon~ \lambda(\varepsilon)~
\frac{\omega_H^2(\varepsilon)}{\nu^2(\varepsilon,u)}, \\
\nonumber
\lambda(\varepsilon)=\frac{B_0(\varepsilon) m^2(\varepsilon)
\nu^2(\varepsilon,u) \left({\ds -\frac{\partial f_0(\varepsilon)}
{\partial \varepsilon}} \right)}
{\omega_H^2(\varepsilon)+\nu^2(\varepsilon,u)},
\eea

\beq
B_0(\varepsilon)=\frac{1}{2\pi^2\hbar^3} \frac{m(\varepsilon)\delta(u)}
{\nu(\varepsilon,u)p^3(\varepsilon)s_0} \int_0^{2p} dq~
\frac{\beta_e(q)~\hbar\omega_q~W_q \tilde{N}(q,T_p)}{\beta(q)}~q^2.
\eeq

We will calculate the thermomagnetic and thermoelectric quantities
for the following cases:

{\bf i.} Electrons transfer their momentum to impurity ions, and phonons 
are scattered preferably by electrons $k=t$. In this case the thermal 
drag of electrons with phonons is prevails.

{\bf ii.} Electrons and phonons are scattered preferably by each other. 
In this case, the mutual drag of electrons and phonons is important 
($k=t, {\ds \frac{\beta_{pb}}{\beta} \gg \frac{\nu_i(\varepsilon)}
{\nu_0(\varepsilon,u)}}$). In the case of heating of electrons and 
phonons simultaneously under the mutual drag conditions, the energy 
gained from the external field by the electron--phonon system transferred
to environment through the crystal boundaries. It is connected with two 
facts. First, the channel of energy and momentum transfer to reservoir
stands very narrow. Second, it is necessary to fulfill the stationarity 
condition of the solution. If the thermal drag dominates, this energy 
is transferred to LW phonons. Each of the cases mentioned consists of 
two subcases:

{\bf a.} Electrons are heated by the external field, but phonons are 
nonheated and have the temperature of lattice, $T_p=T$.

{\bf b.} Both electrons and phonons are heated, $T_e=T_p(E,H)$. 

We consider the NE effect by taking into account the parts of the NE 
voltage which are related to the interaction of LW phonons with electrons
and LW phonons with SW phonons separately. We also consider some special 
cases which include the part of the thermopower connected to the 
interaction of heated LW phonons with SW phonons.

The thermopower of cool and heated charge carriers in degenerate and
nondegenerate impure semiconductors near AIT is investigated in detail
earlier\cite{44,45}.

\section{Thermopower and Thermomagnetic Coefficients}
If ${\bf E}\parallel{\bf H}\parallel\hat{y} \perp \nabla T_e \parallel
\hat{z}$, then under the condition of $\nabla_x T_e=0$ (the isothermal
condition along $x$ axis) taking into account $J_{Tx}=J_{Tz}=0$, from
Eq. (12) for $E_{Tx}$ and $E_{Tz}$ we have

\beq
E_{Tx}=H(Q_e\nabla_zT_e+Q_p\nabla_z T_p),~~~Q_{e,p}=\frac{1}{H}
\frac{\sigma_{11}\beta^{(e,p)}_{12}-\sigma_{12}\beta^{(e,p)}_{11}}
{\sigma_{11}^2+\sigma_{12}^2},
\eeq
where $Q_e$ and $Q_p$ are the electron and phonon parts of the NE 
coefficient (in this work, $E_{Tz}$ is studied only when the magnetic
field is parallel to $\nabla T_z$), respectively. 

At weak magnetic fields, when $\omega_H(\bar{\varepsilon}) \ll 
\nu(\bar{\varepsilon},u)$ from Eqs. (13) and (19) we get

\bea
Q_e=\frac{1}{eH}~\frac{\omega_H}{\nu(\Theta_e,u)}~{\ds \Lambda^{r-2}~
\Theta_e^{2+s(2r-4)}~b_1(s,r)}, \\
\nonumber
Q_p=Q_{p1}(T_e)+Q_{p2}(T,T_e),
\eea
where,

\bea
{\ds Q_{p1}(T_e)=\frac{8}{3}\frac{b_2(s,r)}{eH}\frac{\omega_H}
{\nu(\Theta_e,u)}\frac{\beta_e^0(T)\beta_e(\varepsilon_g,T)}
{\beta^2(\varepsilon_g,T)}~\Lambda^{r}\Theta_e^{2+s(2r-1)}\delta(u)}, \\
\nonumber
{\ds Q_{p2}(T,T_e)=\frac{3~ 2^{h-t}}{(3+h-t)}~Q_{p1}(T_e)~
\frac{\beta_{pb}(T)}{\beta_e(\varepsilon_g,T)}~B^{h-t}
~\Theta_e^{2+s(h-t-1)}~\frac{\nabla_z T}{\nabla_z T_e}}.
\eea

For the scattering of LW phonons by SW phonons $h=1$, for all other cases 
$h=0$; $\omega_H, \nu(\Theta_e,u)$ and $\beta_e^0(T)$ have the meaning of 
usual definitions in the case of parabolic spectrum, $B(T)={\ds \mu s_0
T^{s-1}}$.

\bea
b_1(s,r)=\frac{1}{2^r s^2~\Gamma_1}\left\{\int_0^{\infty} dx~x^{5+s(4r-5)}
~e^{-x}-[3+s(2r-1)]\int_0^{\infty} dx~ x^{4+s(4r-5)}~e^{-x}\right\}, \\
\nonumber
b_2(s,r)=\frac{1}{2^r\Gamma_1} \left\{\Gamma(3+s(4r-1))-\frac{\Gamma_4}
{\Gamma_1}\int_0^{\infty} dx~x^{4+s(4r-5)}~e^{-x} \right\}.
\eea

When $T_p=T$, the term proportional to ${\ds \frac{\nabla_z T}
{\nabla_z T_e}}$ falls out. However, in $Q_{p1}(T_e)$, the term
${\ds \frac{\beta_e^0(T)\beta_e^0(\varepsilon_g,T)}
{\beta^2(\varepsilon_g,T)}}$ must be replaced by ${\ds \frac{\beta_e^0(T)}
{\beta(T)}}$. Besides, for $s=1$, $r=-{\ds \frac{1}{2}}$ and $t=1$, the 
integrals in Eqs. (20) and (21) diverge, and it is necessary to give
either $s<1$ or lower limit of the corresponding integrals for some
variables. For example, from ${\ds x_1=\frac{\varepsilon(q/2)}{T_e}}$. 
In other words, in this case it is necessary to take into account the
inelasticity of the scattering of electrons by LW phonons. 

At high magnetic field, when ${\ds \omega_H^2(\varepsilon) \gg
\nu^2(\varepsilon,u)}$ we have

\bea
Q_e=\frac{2^{r}\zeta^2~ \Gamma_2}{e H~ \Gamma_3}~\frac{\nu(\Theta_e,u)}
{\omega_H}~ \Lambda^{2-r}~\Theta_e^{2s(t-r)-2} \\
\nonumber
\left\{2(r-2s+1)+\frac{\beta_e^2}{\Gamma_2 \beta^2}
\frac{\nu_p(\Theta_e,u)}{\nu(\Theta_e,u)} 
\left[\int_0^{\infty} dx~x^{3s+1} D(x_1) e^{-x}- 
\int_0^{\infty} dx ~x^{3s}D(x_1) e^{-x} \right] \right\},
\eea

\bea
\frac{Q_p}{Q_{p0}(T_e)}=1+\frac{\beta(T)}{\beta_0(T)}
\frac{\Gamma_2}{\Gamma_3}~\Gamma(s(7+h-t)-1) - 
\Gamma(s[11-2r+h-t]-3) \\
\nonumber
+ C_1(\Theta_e,u)+C_2(\Theta_e,u).
\eea
In Eq. (24),

\beq
Q_{p0}(T_e)=\frac{2^{3+r}b_3(s,t)\Lambda^{4-r}}{3\Gamma_3 e H}
\frac{\nu(\Theta_e,u)}{\omega_H}\frac{\beta_e^0(T)
\beta_e(\varepsilon_g,T)}{\beta^2(\varepsilon_g,T)}
\delta(u)\Theta_e^{3(7-2r)-2},
\eeq

\beq
C_1(\Theta_e,u)=\frac{\nu_p(\Theta_e,u)}{b_3(s,r)
\nu(\Theta_e,u)}\left(\int_0^{\infty}x^{7s-2}D(x_1)
e^{-x}dx-\frac{\Gamma(3s-1)}{\Gamma_3} 
\int_0^{\infty} x^{3s}D(x_1)e^{-x} dx \right),
\eeq

\bea
C_2(\Theta_e,u)=\frac{3~ 2^{h-t}}{(3+h-t)}\frac{\beta_p(T)}
{\beta_e(T)}B^{h-t}(T)\Theta_e^{2+s(h-t-1)}\frac{\nu_p(\Theta_e,u)}
{\nu(\Theta_e,u)}\frac{\nabla_z T}{\nabla_z T_e} \\
\nonumber
\left(\int_0^{\infty} x^{7+h-t}D(x_1)e^{-x}dx-\frac{\Gamma(s(7+h-t)-1)}
{\Gamma(3)}\int_0^{\infty} x^{3s}D(x_1)e^{-x} dx \right),
\eea

\beq
b_3(s,r)=\frac{\Gamma_2~\Gamma(3s-1)}{\Gamma_3}-\Gamma(s(11-2r)-3),~~
D_{x1}=\frac{1}{\alpha_4}\int_{\varepsilon(q/2)}^{\infty} 
x^{s(8+t)-4}e^{-x}dx,
\eeq

\beq
\Gamma_1=\Gamma(3+s(2r-1)),~~\Gamma_2=\Gamma(s(7-2r)-1),
~~\Gamma_3=\Gamma(3s+1),~~\Gamma_4=\Gamma(1+s(3+2r)).
\eeq

It is necessary to note that at weak and strong magnetic fields $\Omega 
\ll 1$, but at strong magnetic field ${\ds \frac{\Omega~
\omega_H(\varepsilon)}{\nu(\varepsilon,u)}}$ is of the order of 1. 
Therefore, the terms proportional to ${\ds \frac{\Omega~
\omega_H(\varepsilon)}{\nu(\varepsilon,u)}}$ or ${\ds 
\frac{\nu_p(\Theta_e,u)}{\nu(\Theta_e,u)}}$ are essential under the 
of mutual drag conditions of electrons and phonons, and in these
expressions $r=-t/2$. As it follows from Eqs. (20)--(22) and (23)--(29), 
at weak and high magnetic fields $Q_p=0$ for the case $s={\ds \frac{1}{2}}$ 
and $r=t$ is in accordance with the results of \cite{21}--\cite{23}.

When ${\bf E} \parallel {\bf H} \parallel\nabla T_z \parallel \hat{z}$, 
by taking into account $J_z=0$, $E_{Tz}$ has the form

\beq
E_{Tz}+\frac{1}{e} \nabla_z \zeta(T_e)=\alpha_e \nabla_z T_e
+\alpha_p \nabla_z T_p,~~
\alpha_{e,p}=\frac{\beta_{11}^{(e,p)}+\beta_{13}^{(e,p)}}
{\sigma_{11}+\sigma_{13}},
\eeq
where $\alpha_e$ and $\alpha_p$ are the electron and phonon parts of the 
differential thermopower, respectively.

In order to investigate the thermomagnetic effects by taking into account
the heating of electrons and phonons, and their arbitrary degree of mutual 
drag in the region of drift velocities $0<u<s_0$ or near the acoustical 
instability threshold $(u \rightarrow s_0)$, we must first determine the
dependences of $\Theta_e$ and $u$ on the electric and magnetic fields.   
                 
\section{The Nernst--Ettingshausen Voltage and Integral Thermopower}
The NE voltage $U$ and the integral thermopower $V$, which are interesting
from the experimental point of view, have the form:

\bea
U=-\int_0^{L_x} E_{Tx}dx = U_e+U_p^{LW}+U_p^{SW}, \\
\nonumber
V=\int_0^{L_z} E_{Tz}dz = V_e+V_p^{LW}+V_p^{SW}.
\eea
Here $L_x$ and $L_z$ are the length of the specimen in $x$ and $z$
directions. For arbitrary $r$ and $s$, in the case of heating of LW 
phonons we have:

\bea
V_e=\frac{T}{e}\left\{(2rs-s+3)-\frac{\zeta(T_e)}{T_e}\right\} 
\Theta_e=V_{e0}\Theta_e, \\
\nonumber
V_p^{LW}(T_e)=\frac{8}{3}\frac{\Gamma_4}{\Gamma_1}\frac{T}{e}
\frac{\beta_e^0(T)\beta_e(\varepsilon_g,T)}{(3s+1)
\beta^2(\varepsilon_g,T)}\Lambda^2\delta(u)\Theta_e^{3s+1}=
V_{p0}^{LW}\delta(u)\Theta_e^{3s+1}, \\
\nonumber
V_p^{SW}(T_e)=\frac{3}{3+h-t}\frac{\Gamma(1+(3+2r+h-t))}
{\Gamma_4}\frac{\Delta T}{T} \frac{\beta_p(T)}{\beta(T)}
B^{h-t}(T)V_p^{LW}(T_e)\Theta_e^{2+(h-t+1)}.
\eea
where $\Delta T=T(0)-T(L_z)$. In the absence of heating of LW phonons in 
the expression of $V_p$, the term ${\ds V_p^{SW}}$ is absent and in Eq. 
(32) the term ${\ds \frac{T}{\beta^2(T)} \beta_e^0 \beta_e(\varepsilon_g)
\Theta_e^{3s+1}}$ in ${\ds V_p^{LW}}$ must be replaced by

\beq
\Delta T=\frac{\beta_e^0(T)}{\beta_e(\varepsilon_g,T)}\Theta_e^{3s}.
\eeq

\subsection{The Weak Magnetic Field Case}

{\bf 1.}~~In the case $i$ and subcase $a$, from Eqs. (20), (21) and (32), 
near AIT we have:

\bea
U_e=U_{e0}(T)\Theta_{e1}^{(3-s)/a}, 
~~U_p=U_{p0}(\Delta T)\Theta_{e1}^{2(1+s)/a}\delta(u), \\
\nonumber
V_e=V_{e0}\Theta_{e1}^{1/a},~~V_p=V_{p0}(T)\Theta_{e1}^{3s/a}\delta(u),~~
\Theta_{e1}=\frac{A_1}{\Lambda^{1/2}B^h(T)}\frac{\nu_i(T)}
{\nu_p(s,T)}\frac{\beta_e(T)}{\beta_p(T) \varphi(u)}.
\eea
here $A$, the digital coefficient, is of the order of 1, and $a=2+s(2+h+t)$. 
As it follows from Eq. (34), near AIT under the thermal drag condition and
in the absence of heating of LW phonons, $U_{e,p}$ and $V_{e,p}$ are defined 
with ${\ds\frac{\nu_i(T)}{\nu_p(s,T)}}$ and drag parameter ${\ds \eta=
\frac{\beta}{\beta_{pb}} \gg 1}$. At AIT and at its nearest neighborhood,
both $U_{e,p}$ and $V_{e,p}$ do not depend on${\bf E}$ and ${\bf H}$ or 
$\Theta_e$. Since in the considered case ${\ds u=s_0 \frac{E}{E_1}
\Theta_e^{2-s}}$, the condition $u=s_0=const.$ is realized only when
${\ds \Theta_e=\left(\frac{E_1}{E}\right)^{1/2-s}}$. In other words, as 
$u \rightarrow s_0$, ${\ds \frac{E}{E_1}\Theta_e^{2-s}}$ must tends to 1. 
Here $E_1$ reads

\beq
E_1=\frac{2^{3/2}s^2\Gamma(x_1,4s-1)}{\Gamma(x_1,3s+1)}
\frac{\beta(T)}{\beta_e(T)}\frac{m_n s_0 \nu_i(T)}{e}\Lambda^{1/2}.
\eeq

It is clear from the physical point of view that as $u \rightarrow s_0$, 
the drift velocity of electrons does not depend on the energy of charge
carriers, and $v \sim u \rightarrow const$. Thus

\bea
U_e \sim \left(\frac{\nu_e}{\nu_p}\frac{\eta}{\varphi(u)}
\right)^{3-s/a},~~ 
U_p \sim \left[\left(\frac{\nu_i}{\nu_p}\eta\right)^{2(1+s)}
\varphi^{s(n+t)}(u)\right]^{1/a}, \\
\nonumber
V_e \sim \left(\frac{\nu_i}{\nu_p}\frac{\eta}
{\varphi(u)}\right)^{1/a}, ~~ 
V_p \sim \left[\left(\frac{\nu_i}{\nu_p}\eta\right)^{3s} 
\varphi^{2+s(h+t+1)}(u)\right]^{1/a}.
\eea

Since at AIT point  $u=s_0$, $\varphi(u)$ and $\eta \rightarrow 1$, the
quantities $U_e$, $U_p$, $V_e$ and $V_p$ saturate, and do not depend on
${\bf E}$.

{\bf 2.}~~ In the case $ii$ and subcase $a$, as $u \rightarrow s_0$,
$\Theta_e$ has the form:

\beq
\Theta_e=\left(\frac{A_2}{B^h}\eta \right)^{\frac{1}{2+s(h-1)}},
~~~~~A_2=\frac{\Gamma^2(x_1,1-st)}{\Gamma^2(x_1,4s-1)}
\frac{\Gamma_2}{\Gamma_1}-1.
\eeq

In the case ${\bf E} \parallel{\bf H}$, $u$ has the form

\beq
\frac{u}{s_0}(1-\gamma_0)\delta(u)=\frac{\Gamma^2(x_1,1-st)}
{\Gamma^2(x_1,4s-1)}\frac{\beta_e}{\beta}\frac{e E}{m_n s_0 \nu(s,T)}
\frac{\Theta_e^{2-s(4+t)}}{\Lambda},
\eeq

where $\gamma_0=\gamma(H=0)$. As it follows from Eq. (37), if the coupled 
system of electrons and phonons is scattered by the crystal boundaries, 
$(\beta_p \approx \beta_e)$ for arbitrary dispersion law of electrons, 
they interact with deformation acoustical or piezo acoustical phonons, and 
$\Theta_e \geq 1$. In the case, when the coupled system of electrons and
phonons is scattered by SW phonons $(h=1)$, the piezo acoustical scattering 
mechanism is not limited. Since, for $s={\ds\frac{1}{2}}$, $t=-1$,
$\beta_{pb} \approx \beta_p$, the coefficient $A_2<0$ or $\Theta_e<0$, and 
we have the ``cooling" effect. It is connected with so--called the effect
of ``electrons runaway" near AIT for the case of scattering of electrons
by the nonheated piezo acoustical phonons\cite{43}.

From Eqs. (19), (20), (32), (36) and (38), as $u \rightarrow s_0$ we have:

\bea
U_e \sim \eta^{\frac{3+s(h-1)}{2+s(h-1)}}\frac{E_2}{E},~~U_p=
U_{p0}(\Delta T)\Theta_e^{2-s(1+t)}\sim \eta^{\frac{2-s(1+t)}{2+s(h-1)}},\\
\nonumber
V_e=V_{e0}(T)\Theta_e \sim \eta^{\frac{1}{2+s(h-1)}},~~ 
V_p \sim \eta^{\frac{3s}{2+s(h-1)}}\frac{E}{E_2}.
\eea 
Here 

\beq
E_2=\frac{s\Gamma(x_1,4s-1)}{\Gamma(x_1,1-st)}\frac{\beta(T)}
{\beta_e(T)}\frac{m_n s_0 \nu_p(s,T)}{e}~\Lambda~ B(T).
\eeq

As it is follows from Eq. (39), at weak magnetic fields $U_p$ and $V_e$ do 
not depend on $E$ at AIT point and its neighbourhood; and tends to their 
constant values as $u \rightarrow s_0$ while $V_p$ and $U_e$ depends 
on $E$ linearly and nonlinearly, respectively. Therefore, in the absence 
of phonon heating the mutual drag leads to the dependence of thermomagnetic 
voltages on $E$ and $\eta$ near AIT when $\Theta_e=const.$ in contrast to
thermal drag case. In other words, the mutual drag of electrons and phonons 
but not their heating plays the role of mechanism of regulation of
unlimited grow or fall down of galvano and thermomagnetic effects near AIT.

{\bf 3.} The investigation of the thermomagnetic voltage $U$ and thermopower 
$V$ in the thermal drag case when $T_p=T_e$ denoted that for the drift
velocities $0<u<s_0$, the electronic part of the quantities $U_e$ and $V_e$ 
dominates over the phonon parts $U_p$ and $V_p$. The results obtained for
$0<u<s_0$ show that in some cases $V_p$ and $U_p$ for SW phonons dominate
over that for LW phonons. In fact, for the scattering of phonons by phonons 
$(h=1)$, and scattering of electrons by piezo acoustical phonons $(t=-1)$ 
when $s=1$, $V_p^{SW}$ and $U_p^{SW}$ grow proportionally to $E$, ${\ds
\frac{\nu_e(T)}{\beta_p(T)} \gg 1}$ and ${\ds\frac{m_n v_0^2}{T}\gg 1}$
while $V_p^{LW}$ and $U_p^{LW}$ decrease. Here $v_0={\ds \left(
\frac{\varepsilon_g}{2 m_n}\right)^{1/2}}$. At the nearest neighbourhood 
of AIT point, the thermomagnetic voltages are preferably obtained by the
ratio ${\ds \frac{\nu_i(T)}{\beta_p(T)}\frac{1}{\varphi(u)}}$. In this 
case $V_e$ and $U_e$ decrease as $u \rightarrow s_0$, and reach their 
saturation values at $u=s_0$. In other words, the electronic part of the 
thermopower and thermomagnetic voltages as well as the current saturate 
when $u=s_0$. When $u=s_0$, the stimulated emission and the absorption of 
phonons become equal to each other and the dissipation vanishes. At this
point, electrons emit all the power gained from the external field
spontaneously like phonons, and the stationarity of state is conserved.

{\bf 4.} In the case of $ii$ and subcase $b$ from Eqs. (20), (23) and (24) 
with taking into account Eq. (39), it is easy to obtain that $U_p=const.$,
$V_p \sim \eta{\ds \frac{E}{E_2}}$, and $U_e \sim {\ds\frac{E_2}{\eta E}}$. 
In other words, in contrast to the case of thermal drag in the case of the
mutual drag, the increase of $U$ and $V$ with $\varphi(u)$ near AIT have
a upper limit (in the case of $T_e=T_p$ in Eq. (39), it is necessary to
change the degree of $\Theta$ by $[1-s(4+t)]$). Besides, at $T_e=T_p$ the
scattering mechanism of electrons by piezo acoustical phonons becomes
limited. It means that for the case considered, the heating of phonons
removes the effect of ``electrons runaway".

\section{High Longitudinal $({\bf E}\parallel {\bf H})$ Magnetic Field
Case}

{\bf 1.}~~In the case $i$ and subcase $a$, from Eqs. (24) and (25) with
taking into account $\Theta_e$, from Eq. (34) as $u \rightarrow s_0$ we have:

\begin{equation}
U_p \sim \left(\frac{\nu_i(T)}{\nu_p(s,T)}\eta 
\right)^{\frac{2(2s-1)}{a}}\varphi(u)^{\frac{4+s(h+t-2)}{a}},~~~~~ 
U_e \sim \left(\frac{\nu_i(T)}{\nu_p(s,T)}
\frac{\eta}{\varphi(u)}\right)^{\frac{s-1}{a}},
\end{equation}
$V_e$ and $V_p$ are determined by Eq. (34) as in the weak longitudinal 
magnetic field case. As it follows from Eq. (41) in the case of $s=1$ the
$U_e$ coincides with the expression in the linear theory. In this case 
for ${\ds s>\frac{1}{2}}$, $U_p$ increases proportionally to ${\ds \eta 
\frac{\nu_i}{\nu_p} \eta}$ irrespective of the values $h$ and $t$, and 
it increases with $\varphi$ as $u \rightarrow s_0$. 
 
{\bf 2.}~~As $u \rightarrow s_0$ in the case $ii$ and subcase $a$, from
Eqs. (23) and (24) it follows: 

\begin{equation}
U_e \sim \Theta_e^{s(4+t)-1}\delta(u), ~~~~~ 
U_p \sim \Theta_e^{s(7+t)-2}\delta^2(u).
\end{equation}

By determining $\varphi(u)$ for the case $u \rightarrow s_0$ from 
Eq. (32), and inserting it in Eq. (42), we obtain:

\begin{equation}
U_e \sim \Theta_e^{s(4+t)-1}\delta(u) \sim \eta^{\frac{1}{2+s(h-1)}}
\frac{E}{E_2},~~U_p \sim \eta^{\frac{2-s(1+t)}{2+s(h-1)}}
\left(\frac{E}{E_2}\right)^2.
\end{equation}

From the comparison of Eq.(43) with Eq. (39), in contrast to the weak 
magnetic field case, at $T_e=T$ and in the presence of mutual drag at high
magnetic field both electron and phonon parts of the NE voltage increase
with $E$ and $\eta$ near AIT. In the weak magnetic field case, on the
other hand, under the same conditions $U_e$ decreases with $E$, but $U_p$
tends to saturation, see Eq. (39).

{\bf 3.}~~ At the nearest neighborhood of AIT in the case $i$ and subcase
$b$, we find:

\beq 
U_e \sim \varphi(u)^{\frac{1-s}{3(1+s)}}, ~~~~~
U_p^{LW} \sim \varphi(u)^{\frac{5-s}{3(1+s)}}, ~~~~~
U_p^{SW} \sim \varphi(u)^{\frac{3+s(t-h)}{3(1+s)}}.
\eeq

The thermopower $V$ in this case is obtained from Eq. (39). As it follows
from Eq. (44) 
for $s=1$, $U_e=const.$ and does not depend on $\varphi(u)$, but $U_p$
increases by $\varphi(u)$. When ${\ds s=\frac{1}{2}}$, $U_p=0$ while 
$U_e$ grows as $\varphi(u)$ increases. Thus, at high magnetic field for
the thermal drag case the thermomagnetic quantities do not have upper
limit as $u \rightarrow s_0$  in contrast to the mutual drag case. 

{\bf 4.}~~Consider the case $ii$ and subcase $b$. In this case with taking 
into account $\Theta_e \gg 1$, we have:

\beq
\Theta_e= \left(\frac{3\Gamma(3-3[1+t])\Gamma^2(x_1,4s-1)}
{\Gamma^2(x_1,1-3t)}A(T)\frac{3}{4}\frac{m_ns_0^2}{T}
\frac{\nu_p(s,T)}{\beta_e(\varepsilon_g,T)}\Lambda \right)^{-\frac{1}{st}}.
\eeq

Namely, in the nearest neighborhood of AIT, $\Theta_e=const$. Here, 
${\ds A(T)=\frac{2\pi^2\hbar^3n_0}{(2m_nT)^{3/2}F_0}}$. At AIT, with
due regard for Eqs. (37) and (38) from Eqs. (24) and (25) we have:

\beq
U_e \sim \eta\frac{E}{E_2}, ~~~~~
U_p^{SW} \sim U_p^{LW}\sim \left(\eta\frac{E}{E_2}\right)^2,~~~~~
V_p^{SW} \sim V_p^{LW} \sim \eta\frac{E}{E_2}.
\eeq
Thus, at weak as well as high magnetic fields, the mutual drag of electrons 
and phonons, at AIT point and its nearest neighborhood plays the role
of retaining mechanism of regulation for the unlimited grow or fall down
of the thermomagnetic coefficients as $u \rightarrow s_0$. In the absence
of mutual drag, the thermal drag of electrons by phonons, the phonon part
of the thermomagnetic coefficients in the limit $u \rightarrow s_0$ sharply 
increase as $\varphi(u) \rightarrow \infty$. In the absence of heating of
electrons and LW phonons, their mutual drag leads to the same dependences 
of thermomagnetic effects on ${\bf E}$ and $\eta$ as in Eqs. (39) and (46) 
for the nondiffusion approximation. Finally, we have the effect of
``nonheated nonlinearity" as a result of the mutual drag considered in
\cite{27}. The Eqs. (39) and (46) allow us to have some opinion about the
behavior of electron--phonon system at AIT. The system of electrons and
phonons coupled by the mutual drag (quasiparticle, electron+phonon) heated 
under high external electric field in the condition $u \ll s_0$ at AIT 
point has constant drift velocity $u$, temperature $\Theta_e > 1$ and the
saturation values of kinetic coefficients. In other words, when $u=s_0$, 
electrons emit the power gained from the external electric field like
acoustical phonons and as a result have the stationary state. The state 
$u=s_0$ is a dynamical stationary state. This case is characterized by the 
constant current ($J=const.$) instead of $J=0$ in the ground state.

\section {Thermopower At High ${\bf E} \perp {\bf H}$ Magnetic Field}
The dependences of $V_e$ and $V_p$ on the external electric and magnetic 
in the diffusion approximation, and $V_e$ in the nondiffusion approximation 
are introduced through $\Theta_e$ and $u$. In the case ${\bf E} \parallel 
{\bf H}$, $\Theta_e$ and $u$ do not depend on ${\bf H}$, and, thus, $V_e$ 
and $V_p$ do not consists ${\bf H}$. Only the magnetic field $({\bf E}
\perp {\bf H})$ influences the thermopower considerably through the
dependences on $\Theta_e$ in the diffusion, and $\Theta_e$ and $u$ in the 
nondiffusion approximations. The corresponding coefficients of thermopower
when $({\bf E} \perp {\bf H})$ may be greater significantly than in the 
case $({\bf E} \parallel {\bf H})$. $V_e$ and $V_p^{LW}$ parts of the
thermopower were investigated in detailed in \cite{45}. Here, we consider
the part of the thermopower related to the interaction of LW phonons with
SW phonons, i.e., $V_p^{SW}$.

At high magnetic field $({\bf E} \perp {\bf H})$ irrespective of the 
character of the interaction of electrons by scattering centers, their 
drift velocity ${\ds v=\frac{cE}{H}}$, and, therefore, ${\ds u=
\frac{\beta_e}{\beta}\frac{cE}{H}=\frac{s_0 E}{E_3}}$.

In the case $i$ and subcase $b$ for $0 <u < s_0$, we find $\Theta_e$ as
\beq
\Theta_e=\left[\frac{2^{3/2}\Gamma[x_1,(s-1)]}{3}
\left(\frac{m_ns_0^2}{\varphi(u)}\right)^{1/2}\frac{A(T)}
{\varphi(u)}\frac{\nu_i(T)}{\beta_p(T)}\left(\frac{\beta}
{\beta_e} \right)^2\left(\frac{E}{E_3}\right)^3 \right]^
{\frac{1}{3(1+s)}}.
\eeq

By using Eq. (47) in Eq. (34), we have

\beq
V_p^{SW} \sim \Theta_e^{2+s(l+h+t)}\delta(u)\sim 
\left(\frac{\nu_i(T)}{\beta_p(T)}\right)^{\frac{2+s(2+h-t)}
{3(1+s)}}\left[\left(\frac{s_0 H}{c E}\right)^3 \varphi(u) 
\right]^{\frac{1+s(1+t-h)}{3(1+s)}}.
\eeq

For ${\ds u=\frac{\beta_e}{\beta}\frac{c E}{H}\rightarrow s_0}$, we have

\beq
V_p^{SW} \sim \left[\left(\frac{\nu_i(T)}{\beta_p(T)}\right)^{2+s(2+h-t)}
\varphi(u)^{1+s(1+t-h)} \right]^{\frac{1}{3(1+s)}}.
\eeq

As it follows from (32), for $t=1$ and $h=1$, the dependences of $V_p^{LW}$ 
and $V_p^{SW}$ on $\varphi(u)$ and ${\ds \frac{s_0 H}{c E}}$ are identical, 
but $V_p^{SW}$ increases by ${\ds \frac{\nu_i(T)}{\beta_p(T)}\gg 1}$ faster 
than $V_p^{LW}$. Near AIT, $V_p^{SW}$ does not depend on $\varphi(u)$ and
becomes a constant at $s=1$, $t=-1$, and $h=1$.

If the electrons and phonons are scattered preferably by each other for 
$0< u <s_0$ in the case $ii$ and subcase $b$, we obtain

\beq 
\Theta_e= \left\{s B_1(x_1,t,s)\Gamma_2 A(T)\frac{\beta}{\beta_e}
\frac{\nu_p(s,T)}{\beta_p(T)}\mu T^{s-1}\left(\frac{\beta}{2\beta_e}s_0
-\frac{1}{\varphi(u)}\frac{c E}{H}\right)^{\frac{1}{2-st}}\right\}.
\eeq
Here

\beq
B_1(x_1,t,s)=1-\frac{1}{\Gamma_2}\frac{\beta_e}{\beta}
\int_0^{\infty} x^{3s} D(x_1) e^{-x} dx.
\eeq
By taking into account Eqs. (50) and (34), we have:

\beq
V_p^{SW} \sim \left[\frac{\nu_p(s,T)}{\beta_p(T)}\left(\frac{\beta}
{2\beta_e}s_0-\varphi^{-1}(u)\frac{c E}{H}\right)\right]^{\frac{2+s(r+h-t)}
{3(1+s)}}\left(\frac{s_0 H }{c E}\right)^3 \varphi(u),
\eeq
and near AIT

\beq
V_p^{SW} \sim \left(\frac{\nu_p(s,T)}{\beta_p(T)}\right)^{\frac{2+s(2+h-t)}
{3(1+s)}} \varphi(u).
\eeq

Therefore, both $V_p^{SW}$ and $V_p^{LW}$ increase with ${\ds 
\frac{\nu_p(s,T)}{\beta_p(T)} \gg 1}$ and by $\varphi(u)$ near AIT.

As it follows the foregoing discussion, the thermomagnetic coefficients
near AIT, increase or fall down sharply. But these changes have the 
relative character and have an upper limit (constant value at AIT). These
limitations are connected with the following situations: At AIT, or its 
nearest neighborhood, the equality $u=s_0$ may be satisfied only when 
$\beta=\beta_e$, $\beta=\beta_e+\beta_p$. Therefore as $u \rightarrow 
s_0$, $\beta_{pb} \rightarrow 0$ while $\varphi(u)\beta_{pb}=const$.

\section{Conclusion}
We now give a brief resume of the main results obtained in the present
work. It is shown that in the absence of heating of phonons near AIT,
the scattering mechanism of electrons by piezoelectric potential of
acoustic phonons under the condition of mutual drag of electrons and
phonons is unlimited, i.e., there is effect of so--called ``electrons 
runaway" for this scattering mechanism\cite{43}. It is shown that this
effect is liquidated by taking into account the heating of phonons. At
AIT and its nearest neighbourhood, the cases either the longitudinal 
${\bf E} \parallel {\bf H}$ or the strong ${\bf E}\perp {\bf H}$, 
temperature of electrons is saturated, i.e., it does not depend on $E$
and $H$, and can be obtained from the drag parameter ${\ds \eta=
\frac{\beta(T)}{\beta_p}}$.  

In the case of weak or strong magnetic fields and in the absence or  
presence of phonon heating, the phonon parts of the thermopower and NE
voltage grow infinitely near the point of AIT under the condition of
thermal drag. If the mutual drag of electrons and phonons prevails, then
both in the absence and in the presence of phonon heating the thermoelectric 
quantities near AIT have the same dependence on $E$ as in the case of
absence of heating of electrons and phonons. In other words, at AIT the
electrons and phonons have the same temperature $T_e$ and drift velocity
$u$ (and cools by approaching AIT point). Namely, at the AIT point the
joint temperature $T_e$ and drift velocity (the current ${\bf J}$) of the
system of electrons and phonons coupled by the mutual drag are saturated.
In contrast to the case of thermal drag, in the case of the mutual drag of 
electrons and phonons the thermomagnetic quantities are finite, i.e., 
have upper limit near AIT.

In the absence of phonon heating, the mutual drag leads to the dependence 
of the thermomagnetic coefficients on $E$ and $\eta$ near AIT, when
$\Theta_e=const.$ in contrast to the thermal drag case. In other words,
the mutual drag of electrons and phonons but not their heating plays the
role of regulation mechanism for unlimited grow or fall down of galvano
and thermomagnetic coefficients near AIT, see \cite{27}.

\section*{Acknowledgments}
This work was partially supported by the Scientific and Technical
Research Council of Turkey (TUBITAK). In the course of this work,
T. M. Gassym was supported by a TUBITAK--NATO fellowship.


\newpage
\begin{thebibliography}{99}

\bibitem{1} C. W. J. Beenakker and A. A. M. Staring, Phys. Rev. B
{\bf 46}, 9667 (1992).

\bibitem{2} L. W. Molenkamp, A. A. M. Staring, B. W. Alphenaar and
H. van Houten, {\it Proc. 8th Int. Conf. on Hot Carriers in
Semiconductors}(Oxford, 1993).

\bibitem{3} M. J. Kearney and P. N. Butcher, J. Phys. C {\bf 19},
5429 (1986); {\it ibid.} {\bf 20}, 47 (1987).

\bibitem{4} R. J. Nicholas, J. Phys. C {\bf 18}, L695 (1985).

\bibitem{5} R. Fletcher, J. C. Maan, and G. Weimann, Phys. Rev. B
{\bf 32}, 8477 (1985).

\bibitem{6} R. Fletcher, J. C. Maan, K. Ploog, and G. Weimann,
Phys. Rev. B {\bf 33}, 7122 (1986).
\bibitem{7} D. G. Cantrell and P. N. Butcher, J. Phys. C {\bf 19},
L429 (1986); {\it ibid.} {\bf 20}, 1985 (1987); {\it ibid.} {\bf 20},
1993 (1987).

\bibitem{8} L. D. Hicks and M. S. Dresselhaus, Phys. Rev. B {\bf 47},
12727 (1993).

\bibitem{9} X. Zianni, P. N. Butcher, and M. J. Kearney, Phys. Rev. B
{\bf 49}, 7520 (1994).

\bibitem{10} R. Fletcher, J. J. Harris, C. T. Foxon, M. Tsaousidou,
and P. N. Butcher, Phys. Rev. B {\bf 50}, 14991 (1994).

\bibitem{11} X. L. Lei, J. Phys.: Condens. Matter {\bf 6}, L305 (1994).

\bibitem{12} D. Y. Xing, M. Liu, J. M. Dong, and Z. D. Wang, Phys.
Rev. B {\bf 51}, 2193 (1995).

\bibitem{13} X. L. Lei, J. Cai, and L. M. Xie, Phys. Rev. B
{\bf 38}, 1529 (1988).

\bibitem{14} D. N. Zubarev, {\it Nonequilibrium Statistical
Thermodynamics}, (New York, Consultants Bureau, 1974).

\bibitem{15} E. M. Conwell and J. Zucker, J. Appl. Phys. {\bf 36},
2192 (1995).

\bibitem{16} A. A. Abrikosov, {\it Introduction to the Theory of Normal
Metals: Solid State Physics Suppl}. Vol. 12 (New York, Academic, 1972).

\bibitem{17} B. M. Askerov, {\it Electron Transport Phenomena in
Semiconductors}, (Singapore, World Scientific, 1994).

\bibitem{18} M. Bailyn, Phys. Rev. {\bf 112}, 1587 (1958); {\it ibid}. 
{\bf 157}, 480 (1967).

\bibitem{19} L. E. Gurevich and T. M. Gassymov, Fizika Tverd. Tela
{\bf 9}, 3493 (1967).

\bibitem{20} M. M. Babaev and T. M. Gassymov, Phys. Stat. Solidi(b)
{\bf 84}, 473 (1977).

\bibitem{21} M. M. Babaev and T. M. Gassymov, Fizika Technika
Poluprovodnikov {\bf 14}, 1227 (1980).

\bibitem{22} T. M. Gassymov, A. A. Katanov and  M. M. Babaev,
Phys. Stat. Solidi(b) {\bf 119}, 391 (1983).

\bibitem{23} M. M. Babaev, T. M. Gassymov and A. A. Katanov,
Phys. Stat. Solidi(b) {\bf 125}, 421 (1984).

\bibitem{24} X. L. Lei, C. S. Ting, Phys. Rev. B {\bf 30}, 4809
(1984); {\bf 32}, 1112 (1985).

\bibitem{25} T. H. Geballe and G. W. Hull, Phys. Rev. {\bf 94},
279 (1954); {\it ibid}. {\bf 94}, 283 (1954).

\bibitem{26} M. W. Wu, N. J. M. Horing and H. L. Cui,
cond--mat/9512114.

\bibitem{27} T. M. Gassymov, A. A. Katanov, J. Phys.: Condens.
Matter {\bf 2}, 1977 (1990).

\bibitem{28} T. M. Gassymov, in the book: {\it Nekotorye Voprosy
Eksp. Teor. Fiz.}, (Baku, Elm, 1977), pp. 3--27; Doklady Akademy
Nauk Azerbaijan SSR, Seri. Fiz. Math. Nauk {\bf 32} (6), 19 (1976).

\bibitem{29} T. M. Gassymov, in the book: {\it Nekotorye Voprosy
Teor. Fiz.}, (Baku, Elm, 1990).

\bibitem{30} T. M. Gassymov, Doklady Akademy Nauk Azerbaijan SSR,
Seri. Fiz. Math. Nauk {\bf 32} (6), 3 (1976); T. M. Gassymov and
M. Y. Granowskii, Izv. Akad. Nauk Azerbaijan SSR, Seri. Fiz. Tekh.
Math. Nauk. {\bf 1}, 55 (1976).
\bibitem{31} I. G. Kuleev, I. I. Lyapilin, A. A. Lanchakov,
and I. M. Tsidil'kovskii, Zh. Eksp. Teor. Fiz. {\bf 106}, 1205
(1994) [JETP {\bf 79}, 653 (1994)].

\bibitem{32} I. I. Lyapilin and K. M. Bikkin, in {\it Proceedings
of the 4th Russia Conference on Physics of Semiconductors,
Novosibirsk, 1999}, pp. 52.

\bibitem{33} I. I. Lyapilin and K. M. Bikkin, Fiz. Tekh.
Poluprovodn. (St. Petersburg), {\bf 33} (6), 701 (1999)
[Semiconductors {\bf 33}, 648 (1999)].

\bibitem{34} I. G. Kuleev, A. T. Lonchakov, I. Yu. Arapova and   
G. I. Kuleev, Zh. Eksp. Teor. Fiz. {\bf 114}, 191 (1998) [JETP  
{\bf 87}, 106 (1998)].

\bibitem{35} S. S. Shalyt and, S. A. Aliev, Fiz. Tverd. Tela
{\bf 6} (7), 1979 (1964).

\bibitem{36} S. A. Aliev, L. L. Korenblit, and S. S. Shalyt,
Fiz. Tverd. Tela {\bf 7} (6), 1973 (1965).

\bibitem{37} I. N. Dubrovnaya and Yu. I. Ravich, Fiz. Tverd.
Tela {\bf 8} (5), 1455 (1966).

\bibitem{38} V. I. Tamarchenko, Yu. I. Ravich, L. Ya Morgovskii
{\it et al.}, Fiz. Tverd. Tela {\bf 11} (11), 3506 (1969).

\bibitem{39} K. M. Bikkin, A. T. Lonchakov, and I. I. Lyapilin,
Fiz. Tverd. Tela {\bf 42} (2), 202 (2000) [Phys. Solid State,    
{\bf 42} (2), 207 (2000)].

\bibitem{40} L. E. Gurevich, T. M. Gassymov, Russian Journal of
Solid State Physics (FTP) {\bf 9}, 106, (1967).

\bibitem{41} F. G. Bass, Yu. G. Gurevich, Russian Journal of
Experimental and Theoretical Physics {\bf 52}, 175, (1967). 

\bibitem{Kane} M. M. Babaev, T. M. Gassym, M. Tas, M. Tomak,
submitted for publishing; cond--mat 0107167.

\bibitem{43} T. M. Gassymov, L. E. Gurevich, Fizika Tverd. Tela 
{\bf 11}, 2946, (1969).

\bibitem{44} T. M. Gassymov and A. A. Katanov, Fizika Technika 
Poluprovodnikov {\bf 20}, 1880, (1986).

\bibitem{45} T. M. Gassymov and A. A. Katanov, Fizika Technika
Poluprovodnikov {\bf 22}, 173, (1988).
\end{thebibliography}
\end{document}